# An edge-to-cloud pipeline for near-real-time biological interpretation of echosounder data

Wu-Jung Lee[1,*], Caesar Tuguinay[1,2], Valentina Staneva[3], Lloyd Izard[1], Soham Butala[3], Brandyn M. Lucca[1], Alicia A. Billings[4], Elizabeth M. Phillips[4], John E. Pohl[4], Julia E. Clemons[4]

[1]Applied Physics Laboratory, University of Washington, Seattle, Washington, United States

[2]Department of Electrical and Computer Engineering, University of Washington, Seattle, Washington, United States

[3]eScience Institute, University of Washington, Seattle, Washington, United States

[4]NOAA Northwest Fisheries Science Center, Seattle, Washington, United States

*Corresponding author

Email: leewj@uw.edu

Running head: Edge-to-cloud echosounder data pipeline

## Abstract

Near-real-time information from the ocean is crucial for effective monitoring and decision making. During fisheries and ecological surveys, such information can help timely optimization of survey operations to support assessment of the population status and distribution of ecologically and commercially important marine species. Here, we present an end-to-end, edge-to-cloud pipeline that transforms echosounder data into actionable biological information during acoustic-trawl surveys. Built on Echostack, a suite of open-source Python packages for scalable echosounder data processing, the pipeline operates across the ship ("edge") and the cloud to deliver near-real-time estimates of fish distribution, abundance, and biomass. On the ship, it ingests raw acoustic data, detects target fish presence using a deep learning model, and transmits detection results, reduced acoustic data products, and biometric data from validation trawls to the cloud. In the cloud, it integrates the acoustic and trawl data to estimate target species abundance and biomass and displays the results in interactive web applications. We demonstrated the pipeline during two large-scale fisheries surveys along the west coast of the U.S. in 2025 and 2026, reducing the latency in delivering initial biological estimates from months to minutes. The pipeline's modular design allows easy adaptation to incorporate different acoustic data analysis algorithms and accommodate other survey configurations. It is also readily extensible to an emerging hybrid survey paradigm in which multiple echosounder-equipped uncrewed surface vehicles collaborate with trawl-capable ships to cover broad survey domains. This development represents a step toward incorporating active acoustics into persistent, low-latency ocean observing strategies.

# Introduction

Acoustic-trawl surveys are widely used to estimate the abundance and distribution of ecologically and commercially important species in marine ecosystem studies and fisheries stock assessments (Simmonds and MacLennan, 2005). This survey design integrates acoustic observations of organisms in the water column with targeted biological sampling using trawls deployed from the same vessel. The acoustic backscatter data are collected by high-frequency active sonar systems, or echosounders, which transmit sound pulses and record the returning echoes containing information about organism abundance and distribution. However, acoustic backscatter alone is often ambiguous for determining the taxonomic identity of observed aggregations, and trawl data provide the essential biological context and key biometric parameters such as length, weight, and sex distributions of the ensonified organisms to support specific identification and acoustic inversion. Trawls are spatially discrete and time-consuming, whereas acoustic data are collected continuously, providing high-resolution backscatter measurements across the ensonified area to “connect the dots” between trawls. Together, the complementary acoustic and trawl data enable efficient observation of marine ecosystems at large scale (Godø et al., 2014; McClatchie et al., 2000). Over recent decades, acoustic-trawl surveys have become a standard approach used by scientists and fisheries agencies worldwide to map the distribution and quantify the abundance of many marine species (Benoit-Bird and Lawson, 2016; Fernandes et al., 2002; Simmonds et al., 1992; Stanton, 2012).

The operational landscape of active acoustic ocean observation is rapidly evolving, with data collection expanding beyond traditional research vessels toward greater use of echosounder-equipped autonomous platforms. For example, uncrewed surface vehicles (USVs) are increasingly used to extend the spatial and temporal coverage of acoustic data collection, reduce operational costs associated with large fishery survey vessels, and mitigate unintended fish avoidance of motorized vehicles (e.g., DuFour et al., 2026; Dunn et al., 2023; Handegard et al., 2024; Levine and De Robertis, 2025). These efforts have demonstrated the feasibility of collecting high-quality acoustic data from autonomous platforms, with ongoing work investigating how these platforms can complement traditional ship-based surveys. Because autonomous platforms currently lack trawl capabilities, studies have also explored approaches to integrate acoustic observations from these platforms with biological ground truth from complementary sources to estimate abundance and biomass, with promising early results (Bolser et al., 2023; Chu et al., 2019).

In parallel with these developments, the analysis of echosounder data is also shifting from manual annotation toward automated analysis. The wider availability of broadband echosounders and advances in physics-based scattering models have enabled increasingly sophisticated exploitation of the distinct acoustic scattering properties of different organisms (Andersen et al., 2024; Chu, 2011; Jech et al., 2015; Stanton, 2012; Stanton et al., 1998a, 1998b). As in many scientific disciplines, machine learning (ML) algorithms that learn complex features from acoustic observations are also developing rapidly, driving increasingly automated classification of organisms in large echosounder datasets (e.g., Brautaset et al., 2020; Cote et al., 2025; Dominik Bahlburg et al., 2025; Handegard et al., 2025; Marques et al., 2023; Pala et al., 2024; Zhang et al., 2024). The exponential growth in the volume of acoustic data further drives the need for automated processing to reduce analysis latency and enable more timely extraction of ecological and biological insights from the data.

Despite these advances, an important capability remains largely under-developed in fisheries acoustics: an end-to-end automated pipeline that runs “on-the-fly” during surveys to produce biological interpretations of acoustic observations in near real time. In the broader oceanography domain, near-real-time processing has enabled adaptive sampling for more than a decade. For example, the Software Package for an Adaptive Satellite-Based Sampling for Oceanographic cruises (SPASSO) automatically integrates near-real-time satellite observations and derives Eulerian and Lagrangian diagnostics (Rousselet et al., 2025). These products can be used alongside in situ observations to help identify, monitor, and sample small, ephemeral oceanographic features. Bringing such near-real-time capabilities to active acoustic ocean sensing could transform survey operations by enabling rapid, adaptive decision-making and supporting timely evaluation of sensor configuration and performance. Such information will allow scientists to optimize the allocation of survey resources, including where and when to conduct acoustic sampling and deploy trawls (Thomas et al., 2024). More broadly, sharing near-real-time information across multiple platforms (e.g., ships and USVs) can enable coordinated operations and improved situational awareness across a distributed observing system.

In this paper, we present an end-to-end, edge-to-cloud pipeline designed to deliver near-real-time biological estimates based on echosounder data during ongoing surveys. The pipeline operates across computing infrastructure on the ship (“edge”) and in the cloud to transform raw acoustic data and validation trawl data into spatially resolved estimates of fish abundance and biomass, and was successfully deployed during multiple fisheries acoustic-trawl surveys conducted in 2024-2026. Its modular design allows easy incorporation of alternative analysis algorithms for different survey targets, and is readily extensible to an emerging survey paradigm, in which multiple echosounder-equipped USVs collaborate with trawl-capable ships to conduct efficient, comprehensive surveys. This paper presents the pipeline design in the practical context of field operations, alongside code implementation in the open-source Python software package, `Echodataflow`, to promote widespread adoption and community adaptation.

# Materials and Procedures

## Pipeline overview

The pipeline consists of two primary stages that are typical for processing and interpreting echosounder data to provide biological estimates for downstream ecological or fisheries analyses. Stage 1 involves processing raw echosounder data files and classifying the sources of received echo energy, such as distinguishing different fish and zooplankton species. Stage 2 involves estimating biological quantities, such as abundance and biomass, by combining echo observations with trawl data (Simmonds and MacLennan, 2005). Depending on the survey design and the complexity of biological assemblages in the studied region, information from trawls conducted either concurrently on acoustically detected schools or in close proximity to acoustic survey transects provide important biological “ground truth” for echo classification in Stage 1 and biological estimation in Stage 2. In the implementation presented here that was deployed on the Integrated West Coast Pelagics Survey (IWCPS) between June and September 2025 and 2026, an echogram ML model performed binary segmentation to classify echo sources as Pacific hake (*Merluccius productus*) or not-hake (Lee et al., 2024c). In the context of this paper, which focuses on the overall pipeline architecture and workflow components, this model is an example backscatter classification tool and can be replaced by other rule-based or ML algorithms.

We specifically designed the pipeline to carry out Stage 1 processing on the edge (ship) and Stage 2 processing in the cloud (Fig. 1, Table 1). At the end of Stage 1, intermediate data products are uploaded to cloud storage at user-specified intervals. These include the nautical areal scattering coefficient (NASC, unit: $m^2 \ nmi^{-2}$), an acoustic proxy for biomass, and the mean volume backscattering strength (MVBS or $\overline{S_V}$, unit: dB re 1 $m^{-1}$) averaged over a coarser time-depth grid than the raw data. Separately, biological data from trawls, including the length, weight, and sex of the measured organisms, are also uploaded when new data become available following each trawl. In Stage 2, the pipeline integrates the uploaded acoustic and trawl biological data in the cloud to generate abundance and biomass estimates for the target species, which are displayed on cloud-hosted interactive web applications that are easily accessible via browsers. Other practical operational considerations and detailed processing steps are discussed in the sections below. The pipeline was implemented in the `Echodataflow` package, which leverages functionalities of the open-source Echostack libraries for echosounder data processing (Lee et al., 2024b) and `Prefect` to streamline workflow execution and provide observability (Prefect Technologies, Inc., n.d.). The code implementation is modular, such that individual operational steps of the pipeline are bundled into "flows" that can be orchestrated and monitored through `Prefect` (Fig. 2; also see ***Computational implementation***).

Importantly, this edge-cloud workflow separation decouples the acoustic and trawl data streams until they are combined in the cloud. The pipeline can therefore still function smoothly even when these data flow from different platforms with complementary sampling capabilities, such as when echosounder-equipped USVs operate alongside traditional crewed vessels performing in situ trawl sampling to complete survey missions. This design consideration is expanded in ***Discussion***.

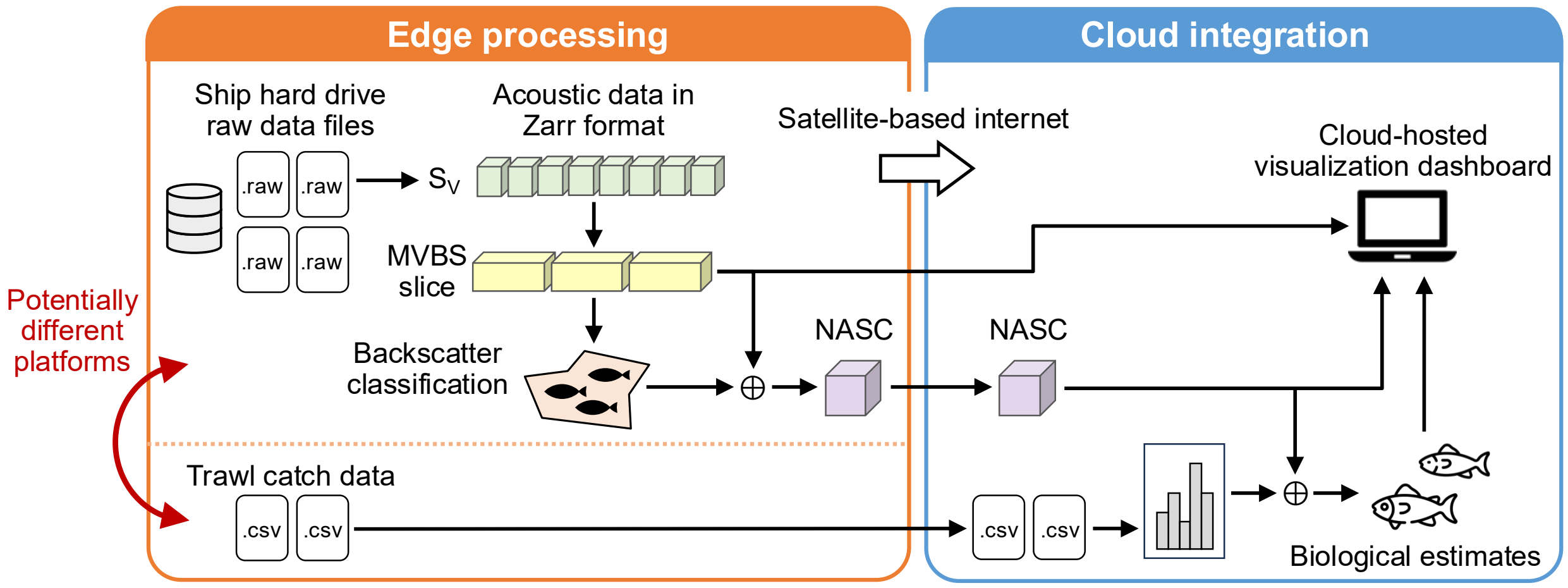

**Figure 1.** Pipeline overview. The pipeline was designed to decouple the edge acoustic and trawl biological data streams until they are combined in the cloud. This design maps directly to an emerging survey paradigm in which USVs (typically collecting only acoustic data) and crewed vessels (often collecting both acoustic and trawl data) operate collaboratively over broad spatial domains to complete the survey mission.

**Table 1.** Summary of edge and cloud flows deployed during IWCPS 2025/2026 (see ***At-sea deployments***). In this table, "flow function" specifies the exact function name used in the `Echodataflow` package, and "schedule interval" indicates intervals of workflow execution or visualization refresh.

| | **Flow operation** | **Flow function or script name** | **Scheduled interval** |
|---|---|---|---|
| **Edge (ship)** | Raw data conversion | `flow_raw2Sv` | 5 min |
| | $S_V$ regridding | `flow_create_MVBS` | 20 min |
| | Backscatter classification | `flow_classify_backscatter` | 20 min |
| | Acoustic data upload | `flow_file_upload` | 5 min |
| | Trawl data upload | `flow_file_upload` | 5 min |
| **Cloud** | Trawl data ingestion | `flow_ingest_haul` | 5 min |
| | NASC data ingestion | `flow_ingest_NASC` | 5 min |
| | Gridded biological estimates update | `flow_update_grid` | Triggered by trawl or NASC data ingestion |
| | MVBS data consolidation | `flow_update_cache_MVBS` | 10 min |
| | Echo data visualization service | `viz_echogram` | 10 min |
| | Biological data and estimates visualization service | `viz_bio` | 10 min |

## Stage 1: Edge processing of ship acoustic data

Edge workflows

The edge (ship) workflow handles all Stage 1 acoustic data processing and uploads both acoustic data products and tabular trawl biological data to the cloud (Table 1). All edge flows are configured to run at user-defined intervals based on a starting timestamp. When this starting timestamp is left unspecified, the workflow will execute based on the actual time; when this starting timestamp is specified, it represents a historical time point for workflow reruns. Below, we describe the operations under each flow and the corresponding function names in the `Echodataflow` package:

- **Raw data conversion (`flow_raw2Sv`)**: This flow checks for unprocessed instrument-generated raw files (`.raw`) and triggers the sequence of data conversion, calibration, and georeferencing operations on each file to produce corresponding volume backscattering strength ($S_V$, unit: dB re 1 m$^{-1}$) data in the Zarr file format (`.zarr`). Zarr is optimized for object storage and distributed computation (Miles, 2019). The flow maintains a ledger of processed filenames to identify any unprocessed files, which could be newly generated files in an actual deployment, or remaining unprocessed files in workflow reruns.
- **$S_V$ regridding (`flow_create_MVBS`):** This workflow combines one or more $S_V$ Zarr files across a user-defined temporal window length, and computes and stores MVBS data products (hereafter referred to as the "MVBS slices") in Zarr files. In the case when $S_V$ data are not available for part of a designated MVBS slice, the missing entries are filled with `Not a Number` (`NaN`) values.

- **Backscatter classification (`flow_classify_backscatter`):** This flow combines one or more MVBS slices to assemble echogram analysis windows according to a user-defined temporal window length (hereafter referred to as the "analysis slices"), performs backscatter classification to generate echogram annotations ( "regions") containing the target species, and computes the NASC associated with these regions. Here, as is usually done in echogram analysis, the target species regions are represented as "masks," which are pixel-level assignments of the scattering sources. Similar to the $S_V$ regridding workflow, when MVBS data are not available for part of a specified analysis window, the missing entries are filled by `NaN` values.
- **Acoustic data upload (`flow_file_upload`):** This flow uploads a user-defined set of acoustic data products to cloud storage. Only data that do not already exist in the cloud storage are uploaded.
- **Trawl data upload (`flow_file_upload`):** This flow uploads trawl biological data in tabular format (e.g., `.xlsx` or `.csv` files) to cloud storage. The data typically include the time, geographic location, and overall species composition of the catch, as well as length, weight, and sex measurements of the target species. These files are ingested in the cloud to derive biological estimates based on the NASC associated with the target species (see next section). Like the acoustic data upload flow, only data that do not already exist in the cloud storage are uploaded.

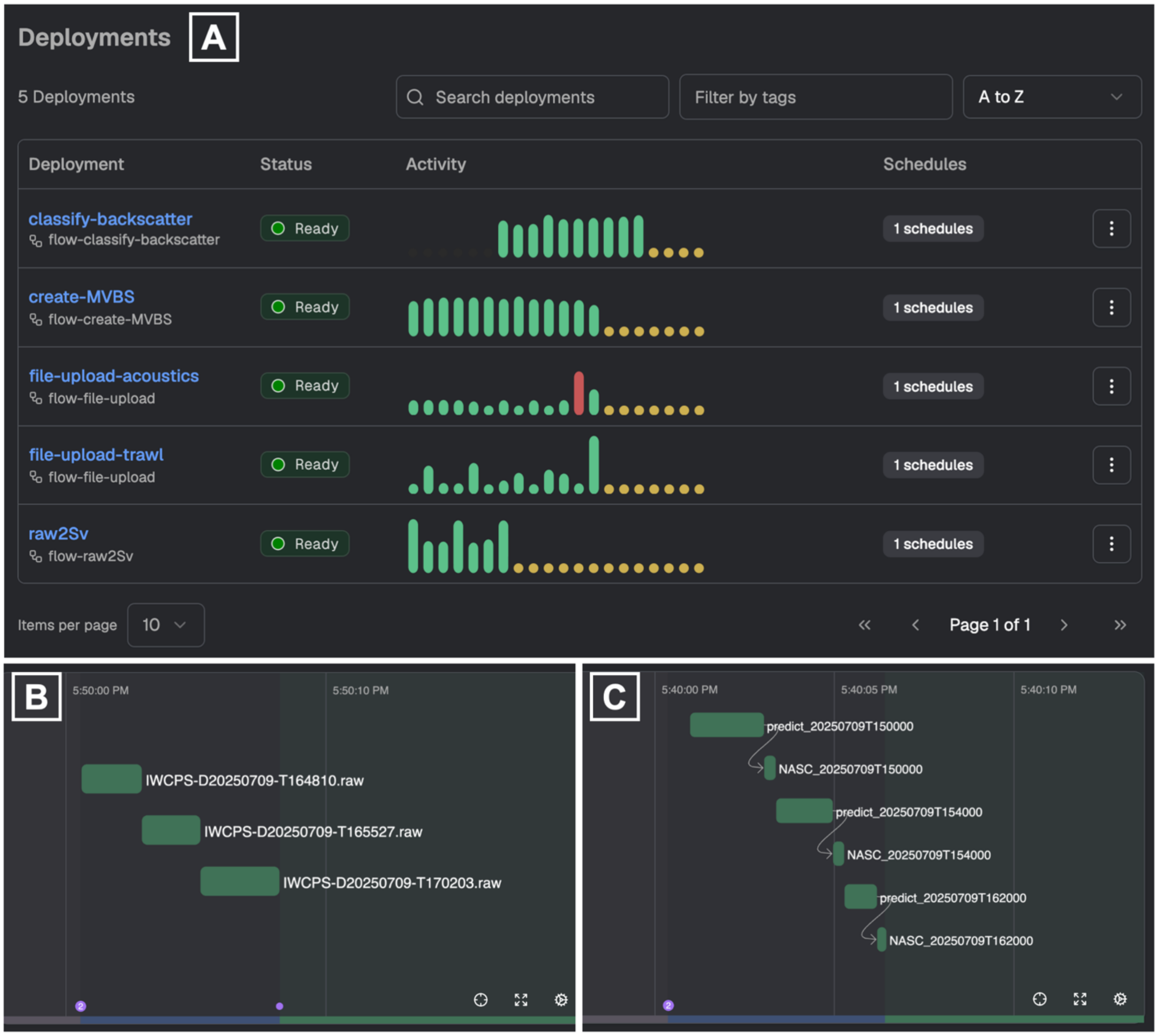

**Figure 2.** The workflow implemented in the `Echodataflow` package leverages functionalities of `Prefect` to provide both overarching and granular observability of workflow execution status. (**A**) Deployment dashboard showing all deployed flows and execution status. Green, red, and yellow bars indicate successful, failed, and scheduled flow executions, respectively. The failed flow execution in this example was due to network connection issues that occurred during acoustic data upload to the cloud; the next scheduled flow run resumed the upload automatically. (**B-C**) Execution details of

flow_raw2Sv and flow_classify_backscatter, which contains individual subcomponents ("tasks") configured with meaningful labels for easy monitoring and diagnosis. For example, tasks under flow_raw2Sv are labeled using the corresponding .raw file names (panel B), and tasks under flow_classify_backscatter are labeled with the starting timestamps of the analysis slices (panel C). Note that the tasks were executed sequentially in this example and do not overlap in time, but they can be configured to execute in parallel to enhance the processing throughput.

### Timing considerations and flow control

The edge flows were executed repeatedly at user-defined intervals (Table 1) aboard a vessel as new data continued to be generated by the echosounder. Specifically, the workflow timing was configured with built-in redundancy to ensure processing robustness. These timing configurations can be tailored to fit the data generation speed, acoustic data analysis methods, and latency tolerance of a given deployment. Below we discuss the timing configurations during the IWCPS 2025/2026 deployments as concrete illustrative examples (Fig. 3).

The raw data conversion flow (flow_raw2Sv) ran every 5 min. During each run, the flow converted all newly available .raw files in the monitored incoming data folder and re-converted the last .raw file processed during the previous workflow execution. Re-converting this last file was necessary, because echosounders typically write data in a streaming fashion by continuously appending to the currently open file until it reaches a user-defined maximum file size. Therefore, during any given raw data conversion flow run, the last `.raw` file typically contained only a portion of the total data it would contain once fully closed (Fig. 3, the rightmost green $S_V$ file). Processing echosounder data as a stream of datagrams instead of discrete files can circumvent this redundancy, but also adds constraints on the deployment network configuration (see ***Discussion***).

An orchestration mechanism was configured to prevent multiple concurrent executions of `flow_raw2Sv` at the same time. This configuration was necessary because at present `flow_raw2Sv` adds converted filenames to the ledger only upon workflow completion. Therefore, if an instance of `flow_raw2Sv` runs for more than the specified execution interval (5 min), another `flow_raw2Sv` run would have been triggered without this orchestration mechanism. During our deployments, `flow_raw2Sv` runs typically finished within 1 min. Extended runs occurred only when there were a large number of unconverted `.raw` files, such as when data collection began long before workflow initialization, when the echosounder collected an unusually large `.raw` file, or when power events interrupted regular edge flow runs for a substantial period of time.

The $S_V$ regridding flow (`flow_create_MVBS`) ran every 20 min to generate three backward-calculated, non-overlapping 20-min MVBS slices, with the end time of the last MVBS slice rounded up to the nearest 20-min boundary following the flow start time. For example, if the flow started at 16:12, `flow_create_MVBS` would compute three MVBS slices with start times 16:20, 16:00, and 15:40. Consequently, the most recent MVBS slice typically did not contain the full 20 min of data (Fig. 3, the rightmost yellow MVBS slice). Under this configuration, each MVBS slice was computed three times. This intentional redundancy ensured that the final set of MVBS slices included all available $S_V$ data from fully closed `.raw` files.

The backscatter classification flow (`flow_classify_backscatter`) similarly ran every 20 min to generate three backward-calculated, non-overlapping 40-min analysis slices, with the end time of the last slice rounded up to the nearest 40-min interval after the flow start time. Like the $S_V$ regridding workflow, the most recent 40-min analysis slice typically did not contain the full 40 min of data (Fig. 3, the rightmost blue analysis window). The analysis slice size depends on the backscatter classification methods used in the workflow. In our deployment, the 40-min length was chosen to match the expected inputs for our ML model (an echogram section of approximately 500 pixels along the `ping_time` dimension, with each pixel spanning 5 sec). During our deployments, the backscatter classification flow occasionally failed when the `ping_time` dimension was too small in the most recent analysis slice for the ML model to function correctly. These errors were corrected automatically at the next flow run when more MVBS data became available in the time window of the last analysis slice.

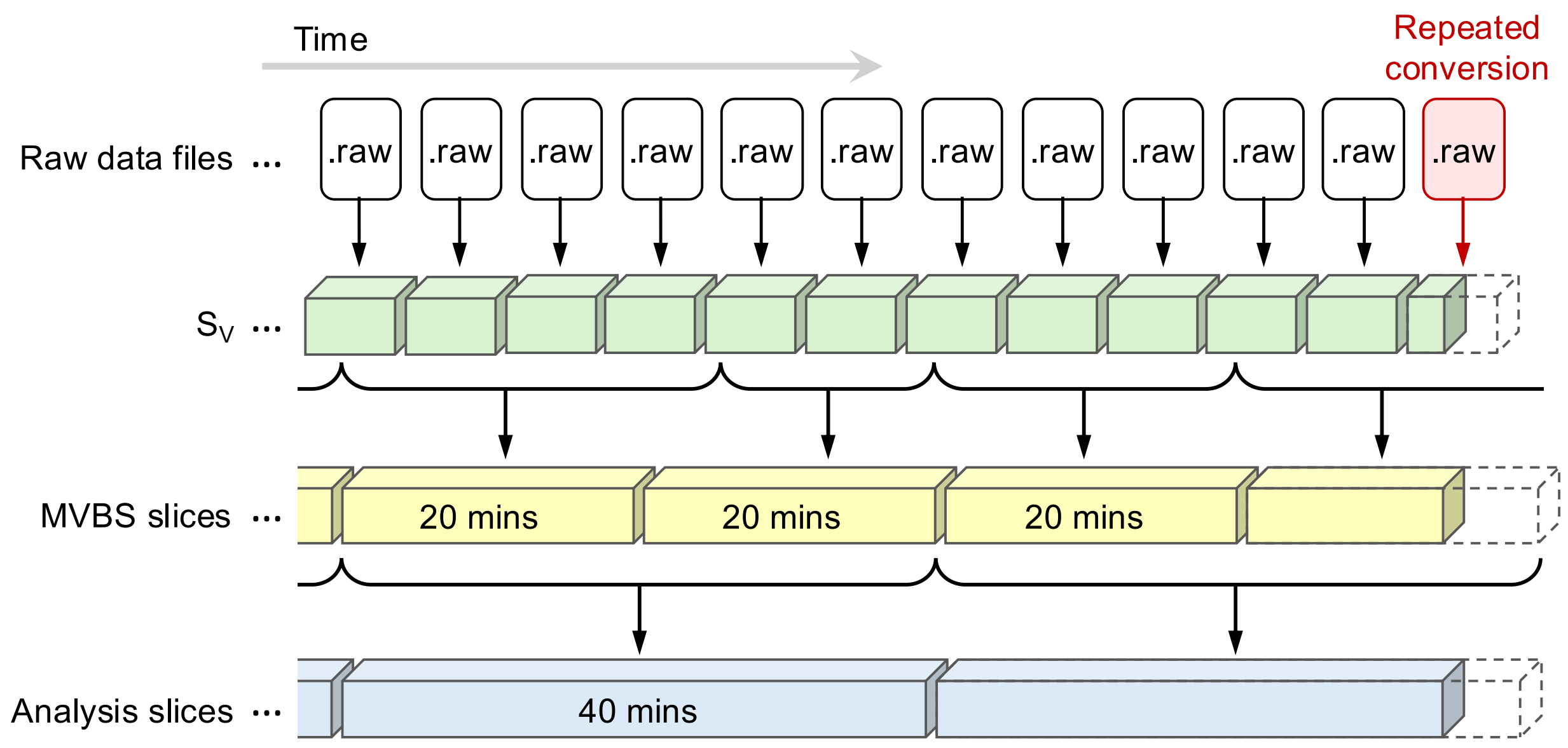

**Figure 3.** Illustration of raw data file processing (`flow_raw2Sv`) and data slicing and combination (`flow_create_MVBS` and `flow_classify_backscatter`). Data from multiple $S_V$ files are sliced and combined to form an MVBS slice. The configured analysis slice length for scatterer classification was 40 min, which happened to align with ping time boundaries of the MVBS slices in this setup. Due to the way the EK80 echosounders write data, the last `.raw` file typically contained only a portion of the total data it would contain once the file was fully closed. Similarly, the last MVBS slice and the last analysis slice typically did not contain the full specified length of data.

## Stage 2: Cloud integration and visualization of acoustic and trawl data

Cloud workflows

The cloud workflow handles all Stage 2 integration of acoustic and trawl data uploaded from the ship via four distinct flows to perform acoustic estimation of fish abundance and biomass, and to consolidate data for interactive web visualization. The operations and corresponding function names in the `Echodataflow` package include:

- **Trawl data ingestion (`flow_ingest_haul`):** This flow checks for newly uploaded trawl biological data, extracts and merges information from each new trawl with all previously consolidated trawl biological data in a tabular format. It then calculates stratified length-weight regression coefficients, mean weight, and mean backscattering cross section ($\overline{\sigma_{bs}}$, unit: $m^2$) of the target species. Downstream workflows use these derived parameters for fish abundance and biomass estimation. During IWCPS 2025/2026, the trawl biological data were stratified using latitude boundaries defined by the International North Pacific Fisheries Commission (INPFC) (Forrester et al., 1983; Phillips et al., 2022).
- **NASC data ingestion (`flow_ingest_NASC`):** This flow checks for newly uploaded NASC Zarr files, calculates acoustics-based biological estimates, including areal number density (unit: number of animals-nmi$^{-2}$), abundance (unit: number of animals), biomass density (unit: kg-nmi$^{-2}$), and biomass (unit: kg) using stratum-specific $\overline{\sigma_{bs}}$ and mean weight from the ingested trawl data, and merges these new data with all previously ingested data in a tabular format.
- **Gridded biological estimates update (`flow_update_grid`):** This flow consolidates all NASC data and biological estimates by computing the mean values of these variables within each grid cell across a set of user-defined spatial grids. This flow provides data for a near-real-time “global” view of the acoustically derived biological estimates via interactive web applications in the cloud (see below). It is triggered by the completion of either the trawl data ingestion or NASC data ingestion workflow, as the grids require updates only when either data source is updated. During IWCPS 2025/2026, each grid cell is 20 nmi × 20 nmi, with land areas excluded (Fig. 4C).

- **MVBS data consolidation for visualization (`flow_update_cache_MVBS`):** This flow combines multiple MVBS slices and extracts a user-specified time window preceding the current time to create an MVBS data cache. This cache supports near-real-time interactive visualization of ship echosounder data in the cloud (see below).

Interactive web applications for data visualization

Two sets of interactive web applications are hosted on a cloud virtual machine (VM) to provide near-real-time visualization of ship echosounder data, detected hake locations, and acoustically derived abundance and biomass information. These applications are run via system services, as they are long-running processes that are best handled outside of `Prefect`. However, note that the visualized data are updated via `Prefect` flow runs (via `flow_update_cache_MVBS`, `flow_ingest_NASC`, and `flow_update_grid`). The interactive web applications refresh periodically (10 min in our implementation; Table 1).

- **Echograms:** Two visualization applications provide near-real-time ship echograms. One application displays the "multi-frequency" echogram of individual frequencies (18, 38, 70, 120, and 200 kHz in IWCPS 2025/2026), and the other app displays the "tricolor" echogram that maps the 120, 38, and 18 kHz data to the R-G-B bands to form a composite image. The tricolor echogram provides a visual summary of the frequency response of scatterers in the ensonified water column and is the input to the echogram segmentation model used in `flow_classify_backscatter` on the edge.
- **Trawl biological data and acoustically derived biological estimates**: Four visualization applications display the results of trawl catches and acoustically derived biological estimates. During IWCPS 2025/2026, these include: (1) stratified histograms of hake fork length grouped by sex; (2) stratified length-weight scatterplots of hake grouped by sex; (3) ship track overlaid with biological estimates associated with hake detections on a bathymetry map (Fig. 4B); and (4) grids of biological estimates of hake on a bathymetry map (Fig. 4C). Users can select among the following displayed biological estimates: NASC, number density, biomass density, abundance, and biomass.

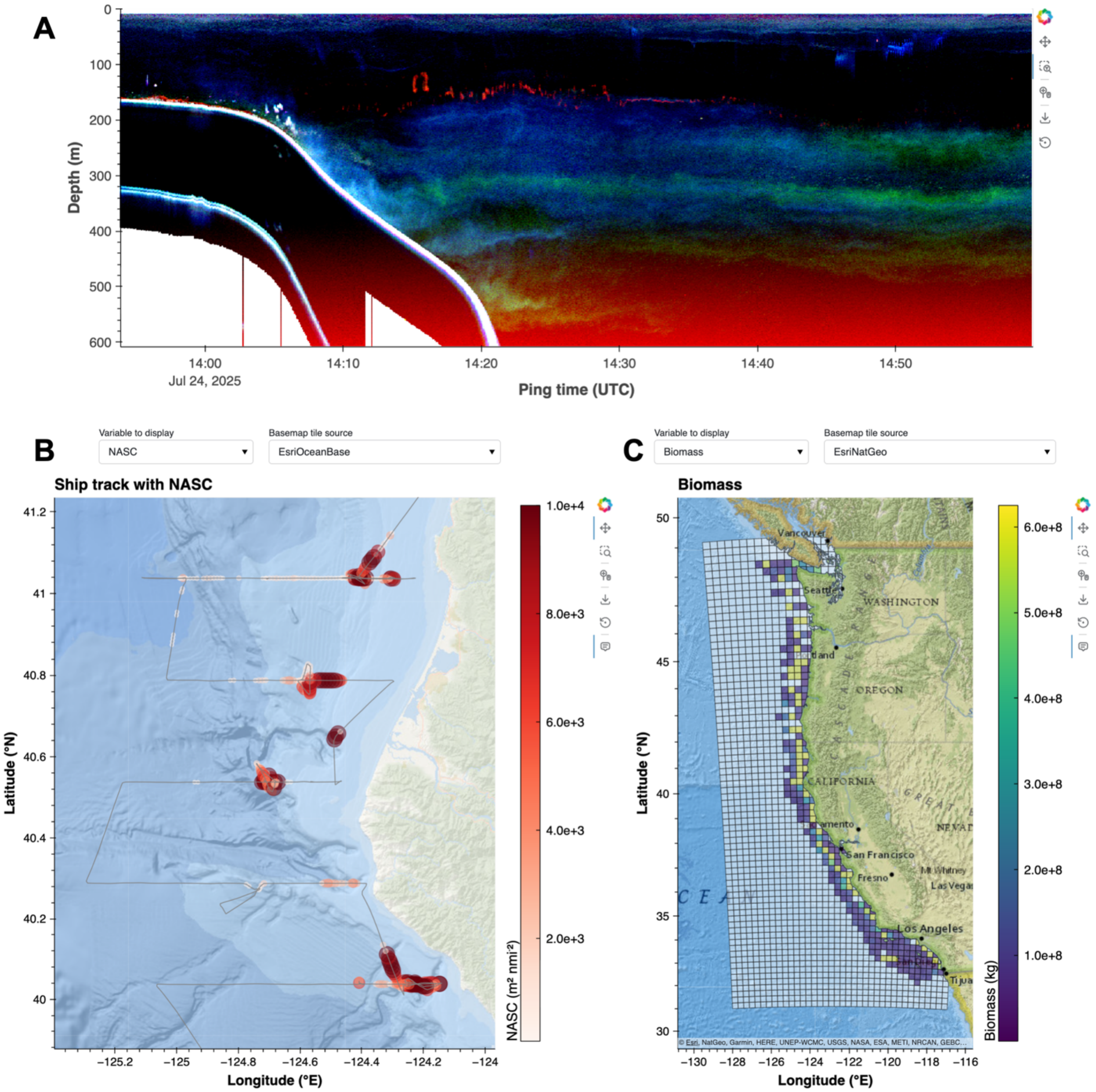

**Figure 4.** Example screenshots of cloud-hosted interactive web applications for visualizing the ship echosounder data and the acoustically derived biological estimates, using IWCPS 2025 data as an example. (**A**) MVBS echogram plotted in the "tricolor" format, mapping the 120, 38, and 18 kHz echo strength to the R-G-B bands to form a composite image. `NaN` values are mapped to white and indicate missing data (e.g., white patches below 400m on the left). (**B**) Target species (Pacific hake) detections

along the survey ship track. Each circle represents the hake NASC value computed in an along-track 0.5 nmi interval. (**C**) Biomass estimates aggregated across a 20 nmi × 20 nmi spatial grid, with land areas excluded.

## Computational implementation

All workflow components are implemented in an open-source Python package `Echodataflow v0.2`, hosted on GitHub (https://github.com/echostack-org/echodataflow). In `Echodataflow`, we leverage the open-source package `Prefect` and system services for robustness and observability. All workflow options are configured by a pair of text-based "recipes" stored as `YAML` files: the deployment recipe (Fig. 5A) controls how the flows interact with the computing resources (e.g., workers and queues) and timing (e.g., execution intervals or trigger settings), whereas the parameters recipe (Fig. 5B) defines the input parameters to the underlying flow functions. Example recipes, including the one presented in this paper, are hosted on a dedicated `echodataflow-recipes` repository (https://github.com/echostack-org/echodataflow-recipes). This recipe-based workflow configuration offers flexibility for iterative experimentation and refinement. `Echodataflow v0.2` represents a major redesign from the earlier v0.1.x implementation, which focused primarily on processing archived data and used different mechanisms for workflow deployment (Butala et al., 2023; Staneva et al., 2024). `Echodataflow v0.2` introduced a new architecture that supports flexible deployment for both archived and near-real-time data processing workflows, and improves code readability, adaptability, and maintainability.

The workflow components extensively use the Echostack libraries for computation and visualization (Lee et al., 2024a, 2024b). The edge flows use `Echopype` to handle echosounder data conversion, calibration, and regridding before feeding organized, analysis-ready data products into the echogram segmentation model for hake detection, and use `Echoregions` to handle hake region masking and contour extraction. The cloud services use `Echoshader` and `HoloViz` [(Yang et al., 2022), especially `Panel`, `HoloViews` and `GeoViews`] to host interactive web visualization applications. Our implementation leverages `Xarray` and `Zarr` functionalities to read, write, and process data smoothly across a local shipboard machine and a cloud VM in a platform-agnostic manner (Bednar and Durant, 2023; Hoyer and Hamman, 2017). We plan to integrate more Stage 2 components into the `Echopop` and `Echoshader` packages in the near future, to further modularize the `Echodataflow` codebase and facilitate wider adoption.

Generative AI was used to assist in: (1) refactoring the `Echodataflow` codebase in April 2026 to reorganize code developed and deployed in 2025 into well-structured subpackages, and (2) the initial implementation of software tests and GitHub Actions workflows to improve the package's continuous integration infrastructure between June and November 2026. The AI assistance was provided through Codex (GPT-5.4, GPT-5.6 Sol, and GPT-6.1 Sol). The authors specified the intended changes and reviewed, modified, and validated the AI-generated code.

**A** Deployment recipe

```yaml
flow_start_time:

source:
  mode: local

default_work_pool_name: edge-local

flows:
  raw2Sv:
    deployment_name: raw2Sv
    deployment_concurrency:
      limit: 1
      collision_strategy: CANCEL_NEW
    interval: 5

  create_MVBS:
    deployment_name: create-MVBS
    interval: 10
    cron_offset: 3

  classify_backscatter:
    deployment_name: classify-backscatter
    interval: 20

  file_upload_acoustics:
    deployment_name: file-upload-acoustics
    flow: file_upload
    interval: 10
    cron_offset: 3

  file_upload_trawl:
    deployment_name: file-upload-trawl
    flow: file_upload
    interval: 10
```

**B** Parameters recipe

```yaml
flows:
  raw2Sv:
    exclude_before: "2025-07-09T12:00:00+00:00"
    exclude_raw_file: []
    parallel: False
    encode_mode: power
    waveform_mode: CW
    depth_offset: 7
    sonar_model: EK80
    nmea_sentence: GGA
    filename_pattern: "IWCPS-*.raw"
    path_raw: /PATH/TO/raw
    path_main: /PATH/TO/local_main
    file_Sv_csv: Sv_files.csv

  create_MVBS:
    slice_mins: 20
    num_slices: 3
    range_bin: 1m
    ping_time_bin: 5s
    path_main: /PATH/TO/local_main
    file_Sv_csv: Sv_files.csv
    file_MVBS_csv: MVBS_files.csv

  classify_backscatter:
    slice_mins: 40
    num_slices: 3
    temperature: 0.5
    softmax_threshold: 0.5
    max_depth: 590.0
    path_weight: /PATH/TO/model_weights/model.ckpt
    path_main: /PATH/TO/local_main
    file_MVBS_csv: MVBS_files.csv
    file_prediction_csv: prediction_files.csv

  file_upload_acoustics:
    src_dir: /PATH/TO/local_main
    dest_dir: ENDPOINT:BUCKET/acoustics
    exclude_subdirs:
      - Sv
      - raw
    max_age: 2

  file_upload_trawl:
    src_dir: /PATH/TO/trawl
    dest_dir: ENDPOINT:BUCKET/trawl
    exclude_subdirs: []
    max_age: 2
```

**Figure 5.** An example pair of the (**A**) deployment recipe and (**B**) parameters recipe for the edge workflow described in this paper. See the `echodataflow-recipes` repository for recipes for the full edge-to-cloud pipeline.

# Assessment

## At-sea deployments

The end-to-end, edge-to-cloud pipeline presented above operated successfully on the NOAA Fisheries Survey Vessel (*FSV*) *Bell M. Shimada* and *FSV Reuben Lasker* during IWCPS 2025 and 2026, respectively, in June-September of each year (NOAA Fisheries, 2025), with near-real-time echograms and derived biological estimates displayed via interactive web applications during each survey. This success built on the experience gained from an earlier prototype tested on the Joint U.S.-Canada Integrated Ecosystem and Pacific Hake Acoustic Trawl Survey (widely known as the "Hake Survey") (Clemons et al., 2024) during the August 2024 research cruise, which allowed us to test and evaluate the performance of several workflow configuration choices regarding flow run schedule, orchestration settings, and server configuration. Operating in the California Current Large Marine Ecosystem, the overarching goals of these surveys are to provide fishery-independent data for stock assessment for Pacific hake and coastal pelagic species, such as Pacific sardine (*Sardinops sagax caerulea*) and Northern anchovy (*Engraulis mordax*).

Separately in 2026, the acoustic data processing and cloud echogram visualization components of the pipeline were also successfully deployed on the NOAA *FSV Oscar Dyson* during the Pollock Acoustic-Trawl survey in the Bering Sea in June-July, 2026. The flows related to backscatter classification and calculating acoustics-based biological estimates were not deployed on *FSV Dyson* due to differences in the ecosystem and survey target species.

During all deployments to date, the acoustic data were collected using Kongsberg EK80 scientific echosounders, with high-level settings summarized in Table 2. Note that while the `Echodataflow` recipes presented here targeted data from EK80 echosounders, the first edge flow (`flow_raw2Sv`, see ***Edge workflows*** and Fig. 5) can be easily adapted to process data from other echosounder instruments supported by `Echopype`.

**Table 2.** Summary of high-level echosounder settings during the 2024-2026 deployments. CW: “continuous wave” narrowband transmission; FM: “frequency-modulated” broadband transmission.

| **Year** | **2024 Hake Survey** | **IWCPS 2025** | **IWCPS 2026** | **2026 Pollock Survey** |
|---|---|---|---|---|
| **Deployment type** | Prototype | Full | Full | Acoustic only |
| **Vessel** | FSV Bell M. Shimada | | FSV Reuben Lasker | FSV Oscar Dyson |
| **Transducers** | 5 channels:<br>18, 38, 70, 120, and 200 kHz | | 6 channels:<br>18, 38, 70, 120, 200, and 333 kHz | |
| **Acoustic data type** | CW, power/angle | Day: CW, power/angle<br>Night: FM, complex | Day: CW, complex<br>Night: FM, complex | CW/FM interleaving, complex |
| **Raw data types handled** | Acoustic and trawl | Acoustic and trawl | Acoustic and trawl | Acoustic only |

## Comparison with conventional workflow

The edge-to-cloud pipeline presented here differs from the conventional human-driven data processing pipeline in two key aspects: (1) it drastically reduces the time required to generate acoustics-based biological estimates of hake using the combination of acoustic and trawl data, and (2) it enables new observational and operational capabilities.

This automated pipeline reduces the time required to generate initial acoustic classifications and associated biological estimates from months to minutes. Conventional echosounder data processing typically proceeds in phases, with the focus shifting from manual echogram annotation during and immediately after the survey to the integration of trawl biological data with acoustic measurements after fieldwork concludes. For IWCPS, the entire process typically takes one to several months of post-survey effort, with key biological estimates only becoming available after all analyses are complete and then provided to stock assessors for further analysis. In contrast, the edge-to-cloud pipeline automatically annotates hake occurrence on echograms as new acoustic data are collected, and continuously updates abundance and biomass estimates as biological data become available following each trawl. Although ML-based results may differ from expert manual analyses (Handegard et al., 2025; Possenti et al., 2026), these preliminary estimates provide valuable information for timely decision making, well before the results of conventional post-survey processing are finalized.

The cloud-based, near-real-time visualization of echosounder data and analysis results (Fig. 4) also provides situational awareness that can drive new operational capabilities and enhance data quality. For example, immediate access to these echograms (Fig. 4A) facilitates collaborative investigation and timely discussions between researchers onboard and onshore to guide operational priorities, such as selecting locations for trawl sampling (Thomas et al., 2024). Early detection of abnormally noisy or corrupted acoustic data, potentially caused by instrument malfunctions or incorrect configurations, enables prompt corrective action during the survey to maintain data quality. Furthermore, mapping target species occurrence along the entire coast (Fig. 4B-C) provides a holistic view of their distributions. In an emerging survey paradigm where multiple crewed and uncrewed vehicles operate collaboratively, such a broadscale view could support adaptive sampling strategies that coordinate vehicle movements to optimize data collection and reduce uncertainty in biological estimates.

## Handling intermediate data products across pipeline steps

For this pipeline that handles real-time instrument-generated raw data, robust workflow execution was achieved by decoupling each flow from the execution status of the preceding flow and instead relying on the availability of its desired input data products. For example, `flow_create_MVBS` and `flow_classify_backscatter` use intermediate data products (Zarr files) generated by `flow_raw2Sv` and `flow_create_MVBS`, respectively, but their execution does not depend on whether the preceding flow run succeeded. Instead, each flow searches for available data products within a specified time window and combines them into a single virtual entity for computation (Fig. 3). When no data are available in the specified time window, the operation was skipped entirely.

This approach of generating intermediate data products that serve as “snippets” of the entire data stream (i.e., creating 20-min MVBS slices in `flow_create_MVBS`) proved more robust than an earlier attempted configuration, in which separate large Sv and MVBS data stores are repeatedly appended and sliced during flow runs, resulting in frequent failures. See ***Supplemental Information Text S1*** for detail.

Note that the combination of data over multiple Zarr files relies on lazy computation, for which only the data needed for active computation are loaded into memory during run time, ensuring built-in scalability of large datasets. This feature was not critical in our deployment, as the data volume was small for each flow run (typical combined volume is 400-700 MB for `flow_create_MVBS` and <20 MB for `flow_classify_backscatter`).

## Orchestration servers and computing resources

In the 2025 and 2026 deployments, a local `Prefect` server hosted on a shipboard computer orchestrated and monitored the edge flows, while `Prefect Cloud` orchestrated and monitored the cloud flows. This setup grew out of our 2024 experimentation, where we attempted to orchestrate both the edge and cloud flows using `Prefect Cloud`. Specifically, orchestrating edge flows from a cloud server with potentially unstable internet connection, which is common for ocean platforms, resulted in frequent failed flow execution: a flow run would either crash or hang indefinitely if internet connection was lost during a run. Therefore, the 2025 and 2026 deployments used a locally hosted `Prefect` server with guaranteed connection to the flow runs, ensuring reliable execution of time-sensitive, near-real-time flows. On the cloud side, for all 2024-2026 deployments, we used `Prefect Cloud` to orchestrate cloud flow runs on a dedicated cloud VM, which also hosted the interactive web applications. Note that it is possible to self-host a server to orchestrate the cloud workflows, and we opted to use `Prefect Cloud` for convenience.

A combination of local filesystem storage and cloud storage supported the ship and cloud flows. All acoustic data products from the edge workflows, including $S_V$, MVBS, segmentation model predictions, and the derived NASC data were stored on a local file system. Except for the raw resolution $S_V$, which were large in volume, all other acoustic data products and the trawl biological data were uploaded to a cloud bucket. The uploaded NASC and trawl biological data were used in the cloud integration workflows, and the uploaded MVBS data were used to generate consolidated echo data for interactive echogram visualization. During our deployments, the MVBS data and backscatter classification results were uploaded to the cloud to allow shore-side evaluation of the ML model performance. Otherwise, only the NASC data corresponding to detected hake aggregations are required for generating biological estimates in the cloud. When network bandwidth or quota constraints exist, MVBS data resolution can be further reduced while still supporting near-real-time echogram visualization. For reference, ***Supplemental Information Table S1*** shows the total size of different data products for each leg of the 2025 survey.

All our deployments ran on commonly accessible hardware and cyberinfrastructure, including a shipboard small form factor desktop computer (a Mac mini with M4 chip, 16GB memory, 512GB internal hard drive, and 4TB external hard drive), a cloud Linux VM with moderate resources (8-core CPU, 30GB memory, 60GB root disk, and 120GB external attached volume), and a cloud bucket for data storage. The cloud resources were provided by the NSF ACCESS program (Boerner et al., 2023) via Jetstream2 (Hancock et al., 2021) and Open Storage Network (OSN) (Kirkpatrick et al., 2021).

## Operational considerations during IWCPS 2025/2026

In this section we discuss aspects of workflow configuration choices driven primarily by operational considerations specific to the IWCPS 2025/2026 deployment.

The near-real-time workflow on the edge had a latency of approximately 20 min relative to echosounder data acquisition. This delay arose because the workflow accessed `.raw` data files hosted on a backup server rather than the primary hard drive to which the echosounder wrote data. This configuration isolated mission-critical survey operations from the processing pipeline. Specifically, the echosounder-generated `.raw` files were synchronized to the backup server at 15 min intervals, while `flow_raw2Sv`, `flow_create_MVBS`, and `flow_classify_backscatter` ran every 5, 20, and 20 min, respectively (Table 1). In addition, we offset `flow_create_MVBS` from `flow_raw2Sv` by 3 min to ensure raw file conversion before MVBS slice creation. Direct access to primary echosounder storage or higher synchronization frequency to the backup server would substantially reduce this ~20 min latency.

Network security restrictions aboard the survey vessels also influenced the deployment configuration. We performed workflow monitoring and configuration through a remote desktop connection [via `Splashtop` (Splashtop Inc., n.d.)]. This approach was acceptable because the workflow computer remained isolated from other mission-critical shipboard systems. For future deployments, remote monitoring and management of the edge flows could follow the principle of least privilege by implementing an authenticated, encrypted tunnel (e.g., through SSH or VPN) that exposes only the local `Prefect` server dashboard, rather than providing remote desktop access with full control of the workflow computer. This approach would also reduce network traffic by minimizing communication overhead.

Finally, the available communication bandwidth dictated the cloud data upload strategy. During the IWCPS 2025 and 2026 field deployments, the edge computer uploaded both regridded MVBS slices (products of `flow_create_MVBS`) and ML hake detection outputs, including the model prediction scores, masks, and NASC (products of `flow_classify_backscatter`), to cloud storage via a Starlink high-speed internet terminal installed on the vessels. This configuration enabled high-resolution cloud-based echogram visualization, as well as near-real-time ML model debugging, experimentation, and assessment during the survey. In deployments with more limited communication bandwidth, the spatial resolution of the MVBS slices along both the `ping_time` and `depth` dimensions can be further reduced, with only the polygon contours of the model-predicted hake regions transmitted alongside the reduced MVBS data.

# Discussion

The primary advance of this work is the development and field deployment of an end-to-end, edge-to-cloud pipeline that transforms raw echosounder data with validation trawl data into acoustics-based biological estimates in near real time during ongoing surveys, with cloud-hosted interactive web visualization applications that make survey information accessible from shore. By closing the latency gap between data acquisition and biological interpretation through a streamlined operational workflow, this pipeline changes when and how survey information becomes available, turning traditionally retrospective data products into an operational capability that can inform survey execution while data collection is still underway. It opens the door for a new mode of survey operation, in which information is shared in near real time across multiple platforms and to operators at different locations (e.g., shipboard, shore-based) to guide adaptive sampling strategies and efficient allocation of personnel and instrument resources.

The pipeline presented here is unique in its integration of edge and cloud operations: Echosounder data are continuously processed on the edge (ship), with intermediate products transmitted to the cloud, where they are integrated with trawl data to generate near-real-time acoustics-based biological estimates. This architecture decouples the acoustic and trawl biological data streams, allowing them to originate from different platforms while still contributing to the same analytical workflow. Such flexibility is particularly valuable as echosounder-equipped USVs expand the spatial and temporal coverage of acoustic data collection, while crewed vessels remain essential for collecting trawl samples that provide the biological ground truth needed to interpret acoustic observations (Chu et al., 2019; Handegard et al., 2024). This pipeline therefore maps directly to an emerging survey paradigm in which USVs and crewed vessels operate collaboratively over broad spatial domains. Moreover, the modular pipeline design readily accommodates additional sensing modalities, such as optical imaging and satellite remote sensing (e.g., Guiet et al., 2026; Molinet et al., 2025; Rasmuson et al., 2022; Scoulding et al., 2023). Such integration can place in situ echosounder observations within a broader oceanographic context by, for example, displaying fronts and other oceanographic features alongside echograms, echo classification results, and vessel tracks. To demonstrate this, a dedicated flow was added during IWCPS 2026 to overlay ship tracks on satellite-derived data products from SPASSO (Rousselet et al., 2025) to relate the acoustic observations with the surrounding oceanographic environment.

The software architecture of our pipeline enables nimble adjustment through the `Echodataflow` package, which follows the "compositional" design philosophy (Bednar and Durant, 2023; Lee et al., 2024b): the workflow is assembled from modular, reusable components that can be selectively combined according to specific mission objectives. This approach contrasts with many existing echosounder software systems that provide real-time or near-real-time data processing on a local machine as a part of a monolithic suite that encompasses data ingestion, organization, analysis, and visualization [e.g., Echoview, Matecho, MOVIES3D, LSSS, etc. ("Echoview," n.d.; Korneliussen et al., 2006; Perrot et al., 2018; Trenkel et al., 2009)]. These systems can be less adaptable when incorporating new functionality or onboarding new developers, as understanding the full system often requires substantial training and cognitive overhead. In contrast, the modular architecture and open source nature of `Echodataflow` offers readability, reusability, and extensibility, making it well suited to the dynamic nature and evolving demands of research and field operations. For example, during the 2026 Pollock Survey (see ***At-sea deployments***), we quickly established an operational remote monitoring system by using only the first two edge flows (`flow_raw2Sv` and `flow_create_MVBS`) and the cloud echogram visualization; in IWCPS 2026, we added a rule-based algorithm for detecting coastal pelagic species on an echogram on the edge alongside other existing flows.

By maintaining both the `Echodataflow` package and curated recipes in `Echodataflow-recipes` (see ***Computational implementation***) as open resources, we aim to provide a collaborative space for reproducible echosounder data workflows. This approach reflects the broader philosophy of `Echostack`: rather than delivering a fixed software product, we aim to enable transparent, community-driven iterations in which researchers and practitioners can collectively improve the tools in response to the changing scientific and operational needs.

One area of future development is to support an alternative raw echosounder data ingestion method that ingests ping-by-ping data as a continuous stream, rather than the current method of ingesting discrete `.raw` files at the file level. We adopted the file-level ingestion approach because, during the 2025-2026 deployments, the workflow could access only a backup server (see ***Operational considerations during IWCPS 2025/2026***) rather than the hard drive to which the echosounder wrote data. This design provides portability, allowing the workflow to operate flexibly without direct connection with critical mission hardware, as long as access to a copy of the `.raw` files is guaranteed. Its primary limitation is a delay of up to the interval at which `flow_raw2Sv` is configured to run (see ***Timing considerations and flow control***). A direct stream of ping-by-ping echosounder data, such as those used by several other software systems (e.g., Echoview Liveview, LSSS, MOVIES3D), could minimize this delay. However, in practice, many analysis methods, including the ML segmentation model we use, operate on short echogram slices or patches rather than individual pings (Brautaset et al., 2020; Ordoñez et al., 2022; Slonimer et al., 2023). In these cases, the additional latency introduced by the current file-based approach can be negligible relative to a streaming implementation.

Another area of future development is to support cloud-to-edge interactions to further enhance deployment flexibility. For example, a “mission file” sent from the cloud to the edge computer could trigger an edge flow that updates the echosounder configuration via REST API supported by the echosounder configuration software. Another possibility is to support human-in-the-loop refinement of the ML model by enabling users to edit model predictions of target species directly in the cloud-based echogram visualization application. The resulting labels can then be used for iterative model retraining (Budd et al., 2021; Handegard et al., 2025). This could substantially improve model performance, as echogram features of biological aggregations vary over time (e.g., across years) and often lead to discrepancies between the training data and previously unseen observations (Brautaset et al., 2020; Handegard et al., 2025).

Separately, changes to the underlying computing environments can further reduce power consumption on the edge and resource demands in the cloud. For example, porting edge flows to a low-power single-board computer or an embedded processor could substantially reduce energy requirements. Implementing cloud flows as serverless functions invoked only when needed, rather than running them on a persistent VM instance, could improve cloud resource utilization. These developments will enhance the efficiency and scalability of future deployments, especially those involving multiple sampling platforms.

The edge-to-cloud pipeline presented here provides a practical framework for integrating and transforming heterogeneous data streams (acoustic and trawl data in this context) into actionable information by combining edge and cloud computing in a modular and scalable manner. This work lays the groundwork for the data and workflow infrastructure needed to incorporate active acoustics into an emerging ocean observing paradigm, in which multiple crewed and autonomous platforms with complementary sensing and sampling capabilities operate collaboratively to enable persistent and adaptive observations of marine ecosystems.

# Conflict of interest

The authors declare no conflicts of interest to disclose.

# Author contribution statement

The author contributions are as follows: **W.-J.L.**: Conceptualization, Data Curation, Funding Acquisition, Investigation, Methodology, Project Administration, Resources, Software, Supervision, Validation, Visualization, Writing – Original Draft Preparation, Writing – Review & Editing. **C.T.**: Data Curation, Software, Validation, Visualization, Writing – Review & Editing. **V.S.**: Conceptualization, Funding Acquisition, Investigation, Methodology, Software, Supervision, Validation, Writing – Review & Editing. **L.I.**: Data Curation, Software, Validation, Visualization, Writing – Review & Editing. **S.B.**: Methodology, Software. **B.L.**: Software, Writing – Review & Editing. **A.A.B.**: Data Curation, Resource, Software, Writing – Review & Editing. **E.M.P.**: Resource, Writing – Review & Editing. **J.E.P.**: Resource, Writing – Review & Editing. **J.E.C.**: Funding Acquisition, Project Administration, Resources, Supervision, Writing – Review & Editing.

# Data availability statement

All raw echosounder data during IWCPS 2025/2026 are available in the NOAA NCEI Water-Column Sonar Data Archive Amazon Web Service S3 bucket: registry.opendata.aws/ncei-wcsd-archive/. All trawl data are available upon request to the NOAA NWFSC Fisheries Engineering and Acoustic Technologies (FEAT) team upon reasonable request.

## Acknowledgments

We thank Alex De Robertis and Scott Furnish at the NOAA Alaska Fisheries Science Center and Elias Capriles and Kevin Stierhoff at the NOAA Southwest Fisheries Science Center for facilitating workflow deployment during the 2026 Pollock Survey and IWCPS 2026, respectively. This work used Jetstream2 at Indiana University and the Open Storage Network through allocations AGR230002 and CIS240371 from the Advanced Cyberinfrastructure Coordination Ecosystem: Services & Support (ACCESS) program, which is supported by U.S. National Science Foundation grants #2138259, #2138286, #2138307, #2137603, and #2138296. This work was supported by NOAA grants #NA20OAR4320271 and #NA24NMFX405C0016 awarded through the Cooperative Institute for Climate, Ocean, and Ecosystem Studies (CICOES), University of Washington.

# Supporting Information

## Text S1

As explained in ***Handling intermediate data products across pipeline steps***, in the current stable implementation, the execution statuses of related flow runs (e.g., `flow_raw2Sv`, `flow_create_MVBS`, `flow_classify_backscatter`) do not impact the execution of the downstream flow. This is achieved by linking these flows only based on the availability of intermediate data products (e.g., $S_V$ and MVBS `.zarr` files), which are each small in size and time span. The downstream flows are configured to use any available intermediate data required for a given flow run and pad `NaN` for any missing data sections.

An alternative approach is to assemble an ever-growing data store by appending newly generated intermediate data, and configure the downstream flow to slice only the required section in each downstream flow run. For example, a single MVBS data store can be assembled by appending new MVBS slices to the existing store in `flow_create_MVBS`, and the downstream flow `flow_classify_backscatter` can then slice the required 40-min MVBS data section to perform echogram segmentation.

We opted for the current approach to disassociate related flows due to two issues we encountered with the alternative approach in the 2024 prototype deployment. First, the Zarr store appending operation can fail intermittently due to decompression errors[1,2]. While Zarr supports parallel tasks to modify independent chunks of a Zarr store, repeated appends to a store that is simultaneously being accessed downstream introduce a more complex synchronization pattern that may require additional safeguards in near-real-time workflows. Second, the appending operation can fail if the new data entries are of different sizes along the `channel` and `depth` dimensions from those in existing Zarr stores. This scenario can occur when the echosounder configuration changes during the survey, such as when the maximum recording range varies depending on the bottom depth. Note that this error occurs only during Zarr store appending. Echopype is capable of handling variable `channel` and `depth` dimensions *within* each `.raw` file.

---

[1] https://github.com/pangeo-data/pangeo/issues/196
[2] https://github.com/pangeo-forge/pangeo-forge-recipes/issues/560

## Table S1

**Table S1.** Total size of acoustic data products generated by the edge workflows in each leg during IWCPS 2025. Only the NASC data is required to generate acoustics-based biological estimates in the cloud. The other data products were uploaded to allow shore-side investigation of the ML model performance. All sizes are in MB.

| | $S_V$ | MVBS | Raw prediction scores | Softmax prediction scores | Prediction masks | NASC |
|---|---|---|---|---|---|---|
| **Leg 1** | 13817.1 | 2704.0 | 514.4 | 290.1 | 2.1 | 7.6 |
| **Leg 2** | 1878.5 | 2540.4 | 475.5 | 276.0 | 2.0 | 7.1 |
| **Leg 3** | 5193.4 | 2670.7 | 536.7 | 309.5 | 2.3 | 8.1 |
| **Leg 4** | 987.3 | 950.5 | 194.6 | 112.3 | 0.8 | 2.9 |
| **Leg 5** | 839.0 | 2497.7 | 500.3 | 286.0 | 2.0 | 7.3 |